\documentclass[aps, twocolumn, pra, showpacs, superscriptaddress]{revtex4}
\usepackage{graphicx}
\usepackage{dcolumn}
\usepackage{bm}
\usepackage{amsmath}

\begin{document}

\title{The lowest scattering state of one-dimensional Bose gas with
attractive interactions}
\author{Yajiang Hao}
\email{haoyj@ustb.edu.cn}
\affiliation{Department of Physics, University of Science and Technology Beijing, Beijing
100083, China}
\author{Hongli Guo}
\affiliation{Institute of Physics, Chinese Academy of Sciences, Beijing 100190, China}
\author{Yunbo Zhang}
\affiliation{Department of Physics and Institute of Theoretical Physics, Shanxi
University, Taiyuan 030006, China}
\author{Shu Chen}
\email{schen@iphy.ac.cn}
\affiliation{Institute of Physics,
Chinese Academy of Sciences, Beijing 100190, China}
\date{\today}

\begin{abstract}
We investigate the lowest scattering state of one-dimensional Bose
gas with attractive interactions trapped in a hard wall trap. By
solving the Bethe ansatz equation numerically we determine the full
energy spectrum and the exact wave function for different attractive
interaction parameters. The resultant density distribution, momentum
distribution, reduced one body density matrix and two body
correlation show that the decreased attractive interaction induces
rich density profiles and specific correlation properties in the
weakly attractive Bose gas.
\end{abstract}


%03.75.Hh Static properties of condensates; thermodynamical,
%statistical, and structural properties
%05.30.Jp Boson systems
%03.75.Kk Dynamic properties of condensates; collective and
%hydrodynamic excitations, superfluid flow

\maketitle
\section{Introduction}

With the rapid experimental progress the ultracold atomic gases have offered
a popular platform to investigate the strongly correlated one dimensional
many-body systems \cite{gorlitz,esslinger,Paredes,Toshiya} for their high
controllability and tunability. One dimensional quantum gases can be
realized with strong anisotropic magnetic trap or two dimensional optical
lattice \cite{Stoeferle,Paredes,Toshiya,HallerPRL} and be described
theoretically by an effective one dimensional (1D) model \cite%
{Olshanii,Olshanii2,Petrov,Dunjko}. In addition the effective 1D interaction
can be tuned from the strongly attractive to the strongly repulsive
interacting regime via the magnetic Feshbach resonance or confinement
induced resonance. Not only the strongly interacting Tonks-Girardeau (TG)
gases \cite{TG} but also two counter-intuitive examples, i.e., the
repulsively bound atom pairs \cite{Winkler} and the super Tonks-Girardeau
(STG) gas-like phase of the attractive atomic gas \cite%
{Haller,Astrakharchik,Batchelor}, have been realized by a sudden
quench of interaction from the strong repulsion to the strong
attraction or vice versa, both of which are hard to realize in the
traditional condensed matter physics and have no analog in solid
state systems. It has been displayed that they are stable excited
states and the stability could be understood from the quench
dynamics of the 1D integrable quantum gas \cite {STGChen}. So far,
the STG gas has attracted intensive theoretical studies from
various aspects \cite
{Kormos,STGChen3,Muth,GirardeauSTG,STGChen2}.

The experimental realization of STG gases and repulsively bound
atom pairs open the door to stable highly excited quantum
many-body phases. This also offers us a method to search for
exotic quantum phases in 1D many-body systems. It has been
predicted that such stable excited states can be prepared in
optical lattice via sudden quantum quench \cite{STGChen3} and the
effective super Tonks-Girardeau gases can be realized via strongly
attractive one-dimensional Fermi gases \cite{STGChen2}.
Theoretically the strongly interacting Bose gases cannot be well
described in the mean-field theory due to the large quantum
fluctuation in 1D system and one has to resort to
non-perturbation methods. Many methods such as Bethe ansatz \cite%
{Dunjko,Fuchs,Hao,Guan}, Bosonization method \cite{Cazalilla}, exact
diagonalization \cite{Deuretzbacher,HaoEPJD}, Bose-Fermi mapping method
(BFM) \cite{Girardeau07} and multi-configuration Hartree theory \cite{Alon}
were used to investigate the 1D quantum gases. It was shown that with the
increase in repulsion strength the ground state density distribution of 1D
Bose gases continuously evolves from a Gaussian-like distribution to a
multi-peak structure while the momentum distribution remains the single peak
structure of bosonic atoms \cite{Hao,HaoEPJD,Deuretzbacher,ferminization}.

By a sudden quench of interaction from strong repulsion to strong
attraction a TG gas shall transfer into a STG gas
\cite{Haller,STGChen,GirardeauSTG}. Because the quantum gas is very
weakly coupled with the environment its energy dissipation is
ignorable and this highly excited state shall be stable. By
decreasing the attractive interaction of STG gas we can investigate
the properties of the lowest scattering state for the Bose gas with
the change of attractive interaction from strongly to weakly
interacting regime. In this work, by numerically solving the Bethe
ansatz equation we obtain exact wave function of the lowest
scattering state of 1D Bose gas trapped in a hard wall potential in
the above interacting regime. We will focus on the density
distribution, reduced one body density matrix (ROBDM) and two body
correlation in the full attractive interaction regime.

The present paper is organized as follows. Section II is devoted to
the description of our model and Bethe Ansatz method. Section III
will give the density profiles, ROBDM and two body correlation in
the full attractive interacting regime. A summary is given in the
last section.

\section{Model and method}

We consider $N$ interacting Bose atoms of mass $m$ confined in a hard wall
box of length $L$, which is described by the Hamiltonian
\begin{equation}
H=-\sum\limits_{i=1}^{N}\frac{\partial ^{2}}{\partial x_{i}^{2}}%
+2c\sum\limits_{1\leq i\leq j\leq N}^{N}\delta (x_{i}-x_{j}).
\label{1}
\end{equation}%
Here the natural unit is used $\hbar =2m=1$ and $c=mg_{1D}/\hbar
^{2}$ is an interaction constant dependent on the effective 1D
interaction strength $g_{1D}$, which can be tuned continuously from
the strong attraction to the strong repulsion by Feshbach resonance
or confinement induced resonance. This model can be solved exactly
by the Bethe ansatz method \cite{OPB}. The many-particle wave
function shall be formulated as the following general form
\begin{eqnarray}
\Psi \left( x_{1},\cdots ,x_{N}\right) &=&\sum_{Q}\theta \left(
x_{q_{N}}-x_{q_{N-1}}\right) \cdots \theta \left( x_{q_{2}}-x_{q_{1}}\right)
\notag \\
&&\times \varphi _{Q}\left( x_{q_{1}},x_{q_{2}},\cdots ,x_{q_{N}}\right)
\label{WF}
\end{eqnarray}%
with
\begin{eqnarray}
&&\varphi _{Q}\left( x_{q_{1}},x_{q_{2}},\cdots ,x_{q_{N}}\right)  \notag \\
&=&\sum_{P,r_{1},\ldots ,r_{N}}\left[ A\left( Q,rP\right) \exp \left(
i\sum_{j}r_{p_{j}}k_{p_{j}}x_{q_{j}}\right) \right] ,  \label{wavefunction}
\end{eqnarray}%
where $Q=(q_{1},q_{2},\cdots ,q_{N})$ and $P=(p_{1},p_{2},\cdots ,p_{N})$
are one of the permutations of $1,\cdots ,N$, respectively, $A\left(
Q,rP\right) $ is the abbreviation of the coefficient $A\left(
q_{1},q_{2},\cdots ,q_{N};r_{p_{1}}p_{1},r_{p_{2}}p_{2},\cdots
,r_{p_{N}}p_{N}\right) $ to be determined self-consistently, and the
summation $\sum_{P}$ ($\sum_{Q}$) is done for all of them. Here $r_{j}=\pm $
indicate that the particles move toward the right or the left, $\theta (x-y)$
is the step function and the parameters $\left\{ k_{j}\right\} $ are known
as quasi-momenta. In the following evaluation the length $L$ will be taken
to be unity unless otherwise specified. For Bosons the wave function should
follow the symmetry of exchange, so the present problem is simplified into
the solution of
\begin{equation}
H\varphi _{1,...,N}\left( x_{1},\cdots ,x_{N}\right) =E\varphi
_{1,...,N}\left( x_{1},\cdots ,x_{N}\right)  \label{sch}
\end{equation}%
in the region of $0\leq x_{1}\leq x_{2}\leq \cdots \leq x_{N}\leq L$ with
the open boundary condition
\begin{equation*}
\varphi _{1,...,N}\left( \cdots ,x_{j}=0,\cdots \right) =\varphi
_{1,...,N}\left( \cdots ,x_{j}=L,\cdots \right) =0.
\end{equation*}%
For simplicity we shall ignore the subscript in $\varphi _{1,...,N}\left(
x_{1},\cdots ,x_{N}\right) $. The wave function in other region $Q$ can be
obtained by the exchange symmetry of Bose wave function.

With some algebraic calculation, the wave function has the following explicit
form
\begin{eqnarray*}
&&\varphi \left( x_{1},x_{2},\cdots ,x_{N}\right) \\
&=&\sum_{P}A_{P}\in _{P}\exp \left[ i\left( \sum_{l<j}^{N-1}\omega
_{p_{j}p_{l}}\right) \right] \exp \left( ik_{p_{N}}L\right) \sin \left(
k_{p_{1}}x_{1}\right) \\
&&\times \prod_{1<j<N}\sin \left( k_{p_{j}}x_{j}-\sum_{l<j}\omega
_{p_{l}p_{j}}\right) \sin \left( k_{p_{N}}\left( L-x_{N}\right) \right)
\end{eqnarray*}%
with
\begin{equation*}
\omega _{ab}=\arctan \frac{c}{k_{b}-k_{a}}+\arctan \frac{c}{k_{b}+k_{a}}
\end{equation*}%
and
\begin{equation*}
A_{p_{1}p_{2}...p_{N}}=\prod_{j<l}^{N}\left( ik_{p_{l}}-ik_{p_{j}}+c\right)
\left( ik_{p_{l}}+ik_{p_{j}}+c\right) .
\end{equation*}%
Here $\in _{P}=\pm 1$ denote sign factors associated with even(odd)
permutations of $P$. The quasi-momenta are determined by numerically solving
the Bethe ansatz equations and the total wave function is given by Eq.(\ref%
{WF}) through $\varphi (x_{1},\cdots ,x_{N})$ under the restriction of
exchange symmetry.

Under the open boundary condition, we can obtain Bethe-ansatz equations
\begin{equation*}
e^{i2k_{j}L}=\prod_{l=1(l\neq j)}^{N}\frac{k_{j}+k_{l}+ic}{k_{j}+k_{l}-ic}%
\frac{k_{j}-k_{l}+ic}{k_{j}-k_{l}-ic},
\end{equation*}%
whose logarithmic forms are formulated as
\begin{equation}
k_{j}L=n_{j}\pi -\sum\limits_{l=1(l\neq j)}^{N}\Big(\arctan \frac{k_{j}+k_{l}%
}{c}+\arctan \frac{k_{j}-k_{l}}{c}\Big).  \label{bae}
\end{equation}%
Here ${n_{j}}$ is a set of integers to determine the eigenstates and for the
ground state $n_{j}=j$ $(1\leq j\leq N)$. The energy of the system is $%
E=\sum_{j=1}^{N}k_{j}^{2}$.

\begin{figure}[tbp]
\includegraphics[height=6cm,width=8cm]{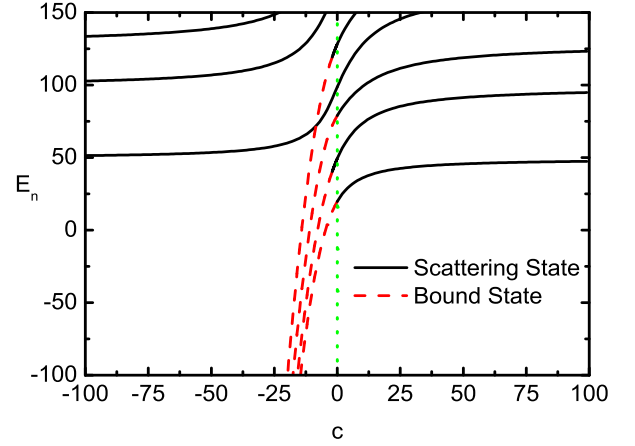}
\caption{(color online) Energy levels of Bose gas with $N=2$ for
different interaction $c$.} \label{fig1}
\end{figure}

For the repulsive interaction ($c>0$) the solutions of Eq. (\ref{bae}) {$%
k_{j}$} are real, while for the attractive interaction ($c<0$) its solutions
might be either real or complex. The real solutions correspond to scattering
states, which can be obtained by solving Eq. (\ref{bae}). We determine the
quantum number {$n_{j}$} in the limit of strong interaction ($c\rightarrow
\pm \infty $) or the limit of zero interaction ($c\rightarrow 0$). In the
former situation the exact wave function can be constructed by means of the
Bose-Fermi mapping method from the wave function of $N$ free Fermions \cite%
{GirardeauSTG}, i.e., the eigenfunction of single particle in the hard wall
trap $sin(j\pi x/L)$ with $j$ being integers. For the ground state $%
j=1,\cdots ,N$ and the excited states are obtained when some $j$ are
replaced by the integers greater than $N$. In the ground state of
noninteracting limit all Bose atoms shall condense into the ground state of
single particle, which gives the exact many body function as $\phi
(x_{1},\cdots ,x_{N})=\prod_{j}sin(\pi x_{j}/L)$. The wave function of
excited states in noninteracting limit, however, takes the form of $\phi
(x_{1},\cdots ,x_{N})=P\prod_{j}sin(n_{j}\pi x_{j}/L)$, where $P$ is an
operator preserving the exchange symmetry of the many body wave function. By
comparing these exact wave function with those obtained from Eq. (\ref{bae}),
the quantum number {$n_{j}$} can be obtained. It is convenient to use the
strong interacting limit here. As $c$ approximates to $\pm \infty $ all
terms in the summation of Eq. (\ref{bae}) vanish such that the quasimomenta $%
k_{j}=n_{j}\pi /L$. Comparing the above exact solution we have $n_{j}=j$ ($%
j=1,\cdots ,N$) for the ground state and the excited states correspond to $%
n_{j}>j$. After deciding the quantum number it is easy to obtain the
solution in the limit of $c\rightarrow \pm 0$. As $c$ approximate $0^{+}$ we
have $k_{j}=(n_{j}-j+1)\pi /L$ and as $c$ approximate $0^{-}$ we have $%
k_{j}=(n_{j}+j-1)\pi /L$.

The complex solutions correspond to bound states, which can be
obtained by assuming the solutions of complex form $k_{j}=\alpha
_{j}+i\Lambda _{j}$ and solving the set of equations of $\alpha
_{j}$ and $\Lambda _{j}$. For example, the complex solutions of
Bethe ansatz equations for the system of $N=2$ are assumed as
$k_{j}=k/2+i\Lambda $ and the Bethe ansatz equations take the
formulation of
\begin{eqnarray*}
kL &=&n\pi +2\arctan {\frac{c}{k},} \\
\exp \left[ 2\Lambda L\right] &=&(-1)^{n}\frac{2\Lambda -c}{2\Lambda +c},
\end{eqnarray*}%
where the integer $n\geq 2$ is quantum number and for the ground
state $n=2$. As an example, we display the full energy spectrum for
$N=2$ in Fig. 1. The scattering states are denoted by solid lines
and the bound states are denoted by dashed lines. It is shown that
the ground state for repulsive case is scattering state and that for
attractive case is bound state. The excited state of attractive Bose
gas in the scattering state has been realized experimentally in Ref.
{\cite{Haller}}. By tuning the 1D interaction constant the Bose
atoms evolve from the weakly interacting Thomas-Fermi regime to the
strongly repulsive TG regime in which the TG gas was realized, and
then by quenching the interaction from the strong repulsion to the
strong attraction the scattering states of attractive Bose gas,
i.e., the STG gas, was realized. Starting from a stable STG gas, we
can further investigate the crossover behavior of the STG gas when
one decreases the attractive interaction very slowly to the very
weak limit. Through an adiabatically slow change of the attractive
interaction, the lowest scattering state of attractive 1D Bose gas
in the whole attractive regime could be reached.

\begin{figure}[tbp]
\includegraphics[width=9cm]{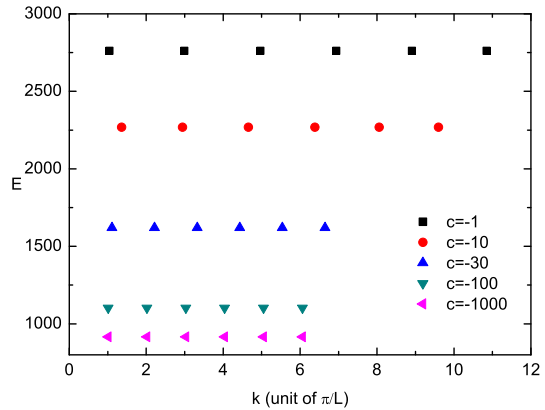}
\caption{ (color online) The quasimomentum distribution and the
corresponding energy for the gas with $N=6$ and $c=-1000,-100,-30,-10,0$
(from below to above).}
\label{fig2}
\end{figure}

In Fig. 2 we display the quasimomentum distribution and the corresponding
energy (longitudinal axis) for the lowest scattering state of the attractive
Bose gases with $N=6$ in the full attractive interacting regime. According
to Eq. (\ref{bae}) in the limit of STG ($c\rightarrow -\infty $) the
solutions of Bethe ansatz equation are equal to those of TG gas ($%
c\rightarrow +\infty $), i.e., $k_{j}=j\pi /L$ $(j=1,\cdots ,N)$. We find
that with the decrease of attractive interaction strength, $k_{j}$ tend to
distribute with larger and larger space between them although the lowest
quasimomentum $k_{1}$ increase first and then decrease. In the limit of
noninteracting limit $c\rightarrow 0^{-}$ we have $k_{j}=(2j-1)\pi /L$ $%
(j=1,\cdots ,N)$ and the space between two neighbor $k_{j}$ has evolved from
$\pi /L$ ($c=-\infty $) to $2\pi /L$ ($c=0$). The energy of the lowest
scattering state increases with the decrease of attractive interaction.

\section{Properties of the Lowest Scattering States for Attractive Bose Gases%
}

In terms of the lowest scattering state wave function $\Psi \left(
x_{1},\cdots ,x_{N}\right) $ the important quantity in one dimensional
interacting many-body system, the ROBDM, can be formulated as
\begin{eqnarray*}
&&\rho (x,x^{\prime }) \\
&=&\frac{N\int_{0}^{L}dx_{2}\cdots dx_{N}\Psi ^{\ast }\left( x,x_{2},\cdots
,x_{N}\right) \Psi \left( x^{\prime },x_{2},\cdots ,x_{N}\right) }{%
\int_{0}^{L}dx_{1}\cdots dx_{N}\left\vert \Psi \left(
x_{1},x_{2},\cdots ,x_{N}\right) \right\vert ^{2}}.
\end{eqnarray*}
The diagonal part of ROBDM gives the expectation values of density
distribution $\rho (x)=\rho (x,x^{\prime })|_{x=x^{\prime }}$, and
the off-diagonal part gives information of coherent properties of
the gas. The Fourier transformation of $\rho (x,x^{\prime }) $ gives
the momentum distribution
\begin{equation}
n\left( k\right) =\frac{1}{2\pi }\int_{0}^{L}dx\int_{0}^{L}dx^{\prime }\rho
(x,x^{\prime })e^{-ik\left( x-x^{\prime }\right) }.
\end{equation}

\begin{figure}[tbp]
\includegraphics[height=8cm,width=8cm]{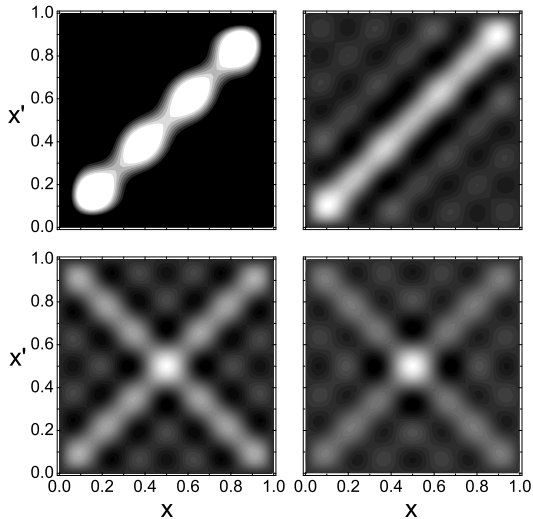}
\caption{The reduced one body density matrix for the gas with $N=4$
for $c=-1000,-10,-1,0$ (from the left above to the right below). }
\label{fig3}
\end{figure}
\begin{figure}[tbp]
\includegraphics[height=6cm,width=8cm]{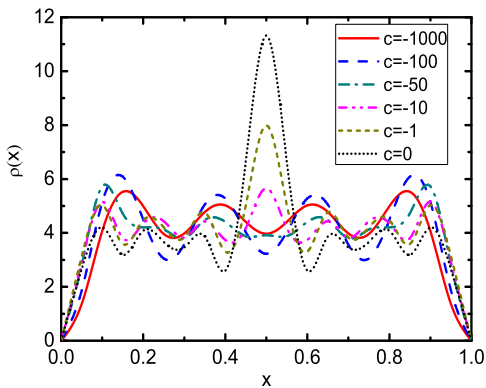}
\caption{(color online) Density distribution for the gas with $N=4$.}
\label{fig4}
\end{figure}

Fig. 3 shows the ROBDM for the lowest scattering state of attractive
Bose gas with $ N=4 $. It deserves to notice that for all
attractively interacting strengthes there exists a strong
enhancement of the diagonal contribution $ \rho (x,x^{\prime })$
along the line $x=x^{\prime }$. In the limit of strong attraction,
the system displays the same behavior as TG gases that $ \rho
(x,x^{\prime })$ reduces rapidly as $|x-x'|$ increases. For weaker
attraction the off-diagonal part of ROBDM shall increase gradually
and we have approximately $\rho (x,L-x) \approx \rho (x,x)$ when the
attraction is weak enough. The density distributions of the lowest
scattering state for different attractive interacting strength are
displayed in Fig. 4. In the strongly attractive limit, it is shown
that the density profile of the STG gas of $N$ atoms exhibits the
Fermi-like shell structure of $N$-peak similar to the density
profile of the TG gas. This is due to the fact that the STG state in
the limit of $c \rightarrow -\infty$ and the TG state in the limit
of $c \rightarrow \infty$ are actually identical. As the attraction
decreases, the density distribution deviates the Fermi-like
distribution and the shell structure oscillates more and more
dramatically. In the limit of $c$ approaching $0^-$ there appear
$2N-1$ peaks in the density profile. The atoms tend to populate at
the center of the trap with the most probability and the density
distribution displays an obvious peak in the center and oscillates
in the region away from the center. In the limit of the strong
attraction atoms populate in the $N$ lowest eigenstates of single
particle so the density profiles show the structure of $N$-peak. As
the decrease of attraction atoms populate at higher eigenstates and
the peak number shall increase. As $c\rightarrow 0^{-}$, atoms
distribute at the $N$ lowest odd states of single particle such that
the density profile exhibits $2N-1$ peaks.

\begin{figure}[tbp]
\includegraphics[height=6cm,width=8cm]{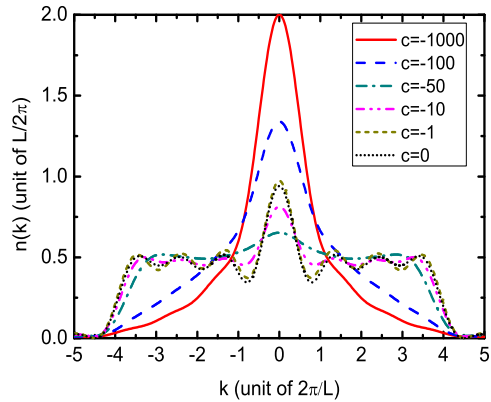}
\caption{(color online) Momentum distribution for the gas with $N=4$. }
\label{fig5}
\end{figure}

The momentum distributions for the lowest scattering state are shown
in Fig. 5. For strong attractive interaction, the atoms accumulate
in the central regime close to zero momentum and the population
distributions decrease rapidly for large momentum, which reflects
the statistics of bosonic atoms. With the decrease of attraction the
atoms distribute more extensively in higher momentum region and in
the strong but finite attractive interaction (e.g. $c=-50$) the
atoms distribute widely in the momentum space without an obvious
zero-momentum peak. For even weaker attractive interaction, the
probability of atoms locating at the zero momentum increases and the
momentum distribution develops a prominent peak at zero point $k=0$.
Away from the central peak, the momentum distribution starts to
display shell structure, which however exhibits distinct feature
from the momentum distribution of Fermi gas. It is because the
lowest scattering state is an excited state that the Bose atoms
populate in higher momentum region.

\begin{figure}[tbp]
\includegraphics[height=8cm,width=8cm]{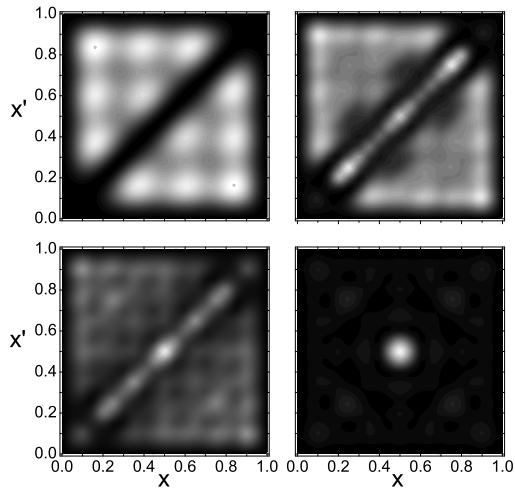}
\caption{Two-body correlation for $c=-1000, -10, -5,0$ (from
the left above to the right below).}
\label{fig6}
\end{figure}

It is also interesting to study the two body correlation function
defined as
\begin{eqnarray*}
&&g_{2}(x,x^{\prime }) \\
&=&\frac{N(N-1)\int_{0}^{L}dx_{3}\cdots dx_{N}\left\vert \Psi \left(
x,x^{\prime },x_{3},\cdots ,x_{N}\right) \right\vert ^{2}}{%
\int_{0}^{L}dx_{1}\cdots dx_{N}\left\vert \Psi \left( x_{1},x_{2},\cdots
,x_{N}\right) \right\vert ^{2}},
\end{eqnarray*}
which denotes the probability that one measurement will find an atom
at the point $x$ and the other one at the point $x^{\prime }$. In
Fig. 6 we display the two body correlation of lowest scattering
state for the Bose gas with $ N=4$. It turns out that two atoms with
strong attractive interaction would try to avoid each other and try
to keep away from each other in certain distance, while as the
decrease in attraction the probability of finding two atoms in the
adjacent region increase and arrive at the maximum as $c\rightarrow
0$.

\section{Summary}

In conclusion we have investigated the lowest scattering state of
Bose gas in the full attractive interacting regime with Bethe ansatz
method. By solving the Bethe ansatz equations numerically the exact
wave functions of the lowest scattering state were determined. Based
on the wave function we obtain the ROBDM, density profile, momentum
distribution and two-body correlation. It is shown that in the STG
limit the ROBDM of the lowest scattering state exhibit the same
behavior as TG gas. With the decrease of attractive interaction the
density distribution evolves from a $N$-peak shell structure to a
$(2N-1)$-peak one and Bose atoms are located at the center of the
trap with the most probability. The momentum distribution manifests
the nature of Bose statistics in the STG limit with an obvious
zero-momentum peak. When the attraction keeps on deviating the STG
limit the momentum distribution spreads more widely although the
most probable position at which the atoms populate is still near the
region of zero momentum. The change of the two body correlation
function with the decrease in the attractive interaction is also
discussed.

\begin{acknowledgments}
This work was supported by NSF of China under Grants No. 11004007, No.
10821403, No. 10974234, and No. 11074153, programs of Chinese Academy of
Sciences, 973 grant No. 2010CB922904, No. 2010CB923103 and National Program
for Basic Research of MOST. Y. Hao was also supported by the Fundamental
Research Funds for the Central Universities NO. 06108019.
\end{acknowledgments}

\end{document}